**Rank-size law, financial inequality indices and gain concentrations by cyclist teams.**

**The case of a multiple stage bicycle race, like Tour de France.**

**Marcel    Ausloos** [1,2,3,*]

[1] School of Business,   College of Social Sciences, Arts, and Humanities, University of Leicester, Leicester,   LE2  1RQ, United Kingdom

email:  ma683@le.ac.uk

[2] Department of Statistics and Econometrics, Bucharest University of Economic Studies, Calea Dorobantilor 15-17, Bucharest, 010552 Sector 1, Romania

email: marcel.ausloos@ase.ro

[3] Group of Researchers for Applications of Physics in Economy and Sociology (GRAPES), Rue de la belle jardinière, 483, Sart Tilman, B-4031 Angleur, Liege, Belgium

 email: marcel.ausloos@ulg.ac.be

# ABSTRACT

This note examines financial distributions to competing teams at the end of the most famous multiple stage professional (male) bicyclist race, "TOUR DE FRANCE". A rank-size law (RSL) is calculated for the team financial gains. The RSL is found to be hyperbolic with a surprisingly simple decay exponent ≈-1. Yet, the financial gain distributions unexpectedly do not obey Pareto principle of factor sparsity. Next, several (8) inequality indices are considered : the Entropy, the Hirschman-Herfindahl, Theil, Pietra-Hoover, Gini, Rosenbluth indices, the Coefficient of Variation and the Concentration Index are calculated for outlining « diversity measures ». The connection between such indices and their « concentration aspects » meanings are presented as support of the RSL findings. The results emphasize that the sum of skills and team strategies are effectively contributing to the financial gains distributions. From theoretical and practical points of view, the findings suggest that one should investigate other "long multiple stage races" and rewarding rules. Indeed, money prize rules coupling to stage difficulty might influence and maybe enhance (or deteriorate) purely sportive aspects in group competitions.

Due to the delay in the peer review process, the 2019 results can be examined. They are discussed in an Appendix ; the value of the exponent (-1.2) is pointed out to mainly originating from the so called « king effect » ; the tail of the RSL rather looks like an exponential.

**Keywords :** Professional cyclist multistage races, Tour de France, Financial gains hierarchy, Financial indices, Rank-size Law

# 1. INTRODUCTION

Income, and more generally wealth, concentration is an old problem in finance and specifically in econometrics (Gini, 1921). Beside the Gini index, concentration ratios measure the percentage of market shares held by firms in some industry (Bikker and Haff, 2002; Marfels, 1972). Those measures have been used in many senses. it is proposed that they can be used in a somewhat generalization scheme in order to check wealth distributions in communities, by analogy with shares of banks on the financial market. This article aims at more specifically contributing to the science of professional sport economics management, in an econophysics perspective, through a specific study, i.e. comparing financial gains of teams in sport competitions, considering how both endogenous and exogenous constraints are influencing outcomes.

On one hand, rankings are ubiquitous processes in human society, leading to defining hierarchies. Their illustrations through rank-size laws (RSLs) are rather common in many cases: sociology, linguistics, lexicography, geography, biology, geophysics, physics, - and economy. Many RSLs look like power laws. The most common one is the Zipf's law (Zipf, 1949). There are many theoretical explanations for its existence. They are related to the Pareto distributions (Newman, 2005). In brief, such explanations are often based on growth or decay processes, occurring in "self-organized systems" (Bak et al., 1988).

On the other hand, empirical relationships between sport competition and economic aspects are parts of the framework in which one measures society's leisures. There is almost no need to recall that ranking is an essential feature of society, the more so in sport competitions, - looking like dynamical self-organizing system (McGarry, et al., 2002), highly tied to "financial rewards" or "money prizes". Managing the prize money rules is an essential duty in order to enhance the competitive aspects, - and the interest of fans as well as that of sponsors and competitors.

This reports aims at discussing ranking aspects of professional cyclist teams in a major (complex) competition, through a financial filter. One is finding an empirical but amazingly simple RSL for the aggregated financial gains of cycling teams in a multiple stage bicycle race, like the "Tour de France". The specificity of the competition and its financial complexity are further emphasized here below. It is shown that the Pareto (20/80) rule is not obeyed in this case, - unexpectedly. Moreover, the calculation of several financial indices serve as support to a discussion; the theoretical and practical connections between such indices are outlined; their meaning is emphasized with respect to such unexpected findings on the rank-size law and concentration aspects. Several indices emphasize the empirical role of the « best teams », others that of the « worst teams ».

Like in for example soccer (Gasparetto & Barajas, 2018 ; Scelles et al., 2013), concentrating on the "best" teams, a « team value » can be measured through prize money which is accrued for the team based on the individual racers (ranking) performances, - in different stages of a long (23 days) race. By investigating data over recent years, about the "Tour de France", it is found that the distributions of aggregated prize money follow an apparently universal power law, with a somewhat unexpected exponent having a « simple » value, nearly equal to -1.

In order to understand the origin of this universal scaling RSL, one can focus on the financial gain distribution rules, and link the prize distribution to the Pareto principle of factor sparsity. However, it is found that the financial gain distributions do not obey the Pareto principle of factor sparsity. Moreover, a structural break is found in 2016.

Beside this so newly found RSL, several (8) financial coefficients, are calculated in the following sections: the concentration coefficient, with the statistical entropy, Gini, Theil, Herfindahl–Hirschman, Rosenbluth, Pietra-Hoover (Ricci-Schutz) indices, and the coefficient of variation; these coefficients are not often encountered, but are of interest for the « diversity and concentration » questions (Allison, 1978), whence are introduced and compared with each other. They serve as a basis of discussion for the findings in the Conclusion section.

For more completely defining the framework, focussing upon « money in sport », one should admit that the literature is of course huge (Neale, 1964; Szymanski, 2003). The same is true about ranking criteria, whence about RSLs (Ausloos and Cerqueti, 2016b; Reed, 2001). For completeness, let us recall that measures of concentration, in particular their significance, have been also discussed at length in many papers (Bikker and Haff, 2002; Marfels, 1972), while inequality indicators were well presented by Foldvary (2006).

However the "numerical intersection" of such sets seems to contain no element, thereby quite shortening a concise literature review on the true state of the art. Thus, in order to save space, and be very specific, the most relevant papers will be only mentioned in the Methodology (Section 2) and Data Analysis (Section 3) with an appropriate comment if necessary.

In Section 4, one provides a discussion of the findings, in particular the breakdown of Pareto law. A set of conclusions is found in Section 5. Due to the delay in the peer review process, the 2019 RSL can be examined. It is displayed and discussed in an Appendix.

## 2. METHODOLOGY

*Specificity of Professional Cyclism Sport Competition*

There are several ways to observe and to study financial concentration and RSLs in "self-organized systems" within the sport competition realm. First of all, one should divide the sport landscape distinguishing "professionals" from "amateurs". The former case is considered here below. Next, one may specify a certain type of sport and observe the activities through national or international sport leagues: sport leagues make teams and individuals mutually dependent for their existence and visibility. Within sport leagues, one can also distinguish different types of business.

On the other hand, one can see sport as individual or team competitions. A very interesting case, from many points of view, is when individual skills merge into a team result (Baumeister et al., 2016). This seems to be the case of almost all team competition in leagues, but holds a very specific feature in cycling races (Forster and Pope, 2004), because only one individual usually wins[2], - in erroneously so called "individual races". In fact, the winner often claims some help from his/her teammates; he/she often shares his/her financial gains with his/her teammates.

Cycling competitions seem different from those in sports like football, rugby, (ice or not) hockey, water polo, etc., in which, from a team point of view, all partners seem to be equal. Salaries and marketing contracts of individuals much differ, of course. However in cycling competitions, the story is quite different[3]: "cycling is an individual sport practiced in team" (Forster and Pope, 2004; Rebeggiani and Tondani, 2006), - "neither a classical team sport, like basketball or football, nor a pure single sport, like tennis". There is a truly planned team strategy or activity in order to favorize one member, who is the most likely intended winner in a race. The 2018 year is in fact very remarkable in that respect: due to such a strategy, more than 70 "individual" races were won by a Quick-Step Floors rider, - 14 different winners out of a 28 men team. The subsequent questions arise: which team is wining the most money? How is the prize money distributed between teams? Can one expect some sort of "Pareto law phenomenon universality" to hold true?

There will not be here any discussion on how much financial gains such a winner gets, or brings in the team common pot, nor whether the reward rules « make sense », but rather, it is interesting to observe how much such a team globally obtains from the variously skilled teammates winning a race, - or receiving some bonus. Such a complex case is found in multistage competitions like the famous (male) bicycling races, Tour de France, Giro d'Italia, Vuelta, and nowadays much copied all over the world, by various organizers.

In order to discuss gain inequality and team hierarchy, one can usefully calculate so called « concentration indices »: the entropy measure, the Theil entropy, the Gini coefficient, the Pietra index, the coefficient of variation, the Rosenbluth (Rosenbluth, 1955; Delalić et al., 2018), and first, the Hirschman-Herfindahl index.

### *Rank-size law, Gains Concentration and Inequality Aspects*

A simple, i.e. power law, "rank-size law" (RSL) can be derived from an (approximated) analytical form presented by the variable, here the gain ($g_i$), ranked in descending order as a function of the (discrete) index *i* giving the "rank" of the team. The "cumulative

concentration distribution curve" (CCDC) is

$$\sum_{i=1}^{N} g_i$$

When the CCDC goes over an 80% threshold, this defines the Pareto rank $r_P$. The Pareto's law expects this rank $r_P$ to be equal to $N/5$.

For completeness, when recalling graphical displays, let it be reminded that the Lorenz curve (LC) originally displays the proportion of income assumed by the % of the people, ranked from the poorest to the richest. By extension, LC is here the gain (in %) of the teams, ranked in increasing order

$$p_i = g_i / \sum_{i=1}^{N} g_i$$

Marfels (1972) distinguishes several types of concentration ratios, according to their weighting schemes and their structure, which can be discrete or cumulative (Bikker and Haff, 2002). Beside such ratios, inequality aspects are often discussed in order to tackle on social aspects (Cowell, 1977). Different types of ratios are considered with different aims as explained below.

A standard measure of (market) concentration (Matsumoto et al., 2012) is the Hirschman-Herfindahl index (HHI).

$$HHI = (\Sigma_i^N g^2_i) / (\Sigma_i^N g_i)^2 \qquad (1)$$

for $N$ agents (teams, here)[4].

Information theory provides another concentration measure, emphasizing the "system disorder", the entropy $S = - p_i \ln (p_i)$, in terms of $p_i$, the proportion of the « size » (financial gain) for the $i$-th team over the whole (gain) ensemble, as previously defined. Notice that $HHI = \Sigma_i^N p_i^2$.

The Theil index Th is a statistical measure of economic inequality (Allison, 1978). Th is tied to the entropy: $Th = (1/N) \Sigma_i^N \hat{g}_i \ln (\hat{g}_i)$, where $\hat{g}_i = g_i /_{<g_i>}$ in which $_{<g_i>}$ is the mean of the gain distribution, i.e. $_{<g_i>} = (1/N) \Sigma_i^N g_i$

The coefficient of variation (CV) is a relative dispersion measure, pointing to the dispersion ($\sigma$) around the mean ($\mu$) of the distribution; $CV = \sigma / \mu$, expressed in percentage, it is somewhat hinting to inequalities. Even though, the skewness and kurtosis lead to a better description of the asymmetry and peakedness of a distribution, whence of inequalities, they are rarely discussed, whence are not further discussed here (Ausloos and Cerqueti, 2018).

Closely related, the Pietra inequality index (Frosini, 2012), also known as the Hoover index,

$$PHI = [\Sigma_i^N ( g_i - <g_i>)] / [2 \ \Sigma_i^N (g_i)] \qquad (2)$$

indicates how the variable values should be distributed in order for them to create a perfect equality or minimal concentration.

An indirect index is the Gini index, Gi, one of the most often used inequality concentration measures. Derived from the LC, it is given by the ratio between the area enclosed by the LC and the diagonal (representing the equality distribution) and the total area below this diagonal.

Related to the Gini index, one has the Rosenbluth index (Rosenbluth, 1955)

$$RI = 1 / [N(1-Gi)]. \qquad (3)$$

"Finally", the so called concentration coefficient (CC) is

$$CC = N/[(N-1) \ Gi]. \qquad (4)$$

## 3. DATA AND NUMERICAL ANALYSIS

The data about the financial gains of the teams in the recent Tour de France years is aggregated from various websites:

http://www.portailduvelo.fr/tour-de-france-2015-les-gains-finaux-par-equipes/

http://www.portailduvelo.fr/tour-de-france-2016-les-gains-definitifs-des-equipes-en-fin-depreuve/

http://www.portailduvelo.fr/tour-de-france-2017-les-gains-des-equipes/

and

http://www.be-celt.com/2018/07/30/tour-de-france-les-gains-des-equipes-le-prix-de-la-souffrance/

For a general information point of view, recall that the (here, $N=22$) competing teams are comprised of originally 9 riders, in the recent years 2018, 2017, 2016, and 2015; at the end of a day stage, money prizes are attributed, according to some pre-established rules, to riders or even to a team; these gains are accumulated. Notice that several teams (18) are imposed by "Union Cycliste Internationale" (UCI) rules and a few (4) are invited by the "Amaury Sport Organisation" (ASO) organizers. The teams (and of course the riders) differ each year.

**PLEASE PLACE FIGURE 1 ABOUT HERE**

It is of course trivial to rank the 22 teams according to their final gains at the end of the competition. One can observe on Fig. 1 that the RSL looks like a hyperbola. The most immediate analysis suggests to obtain the curve characteristics through a fit to a Zipf's law:

$$s = \alpha \; r^{-\gamma} \tag{5}$$

where $\alpha$ and $\gamma$ are parameters to be calibrated for each sample under investigation, and in the present case $s$ is $g_i$. Through a Levenberg-Marquardt algorithm, the best fit to Zipf's law leads to parameter values given in Table 1; the decay exponent $\gamma$ is seen to be close to 1.

**PLEASE PLACE TABLE 1 ABOUT HERE**

The "cumulative concentration distribution curve" (CCDC) of team financial gains in recent Tour de France races is displayed in Fig. 2. From the latter, one obtains $p_i$ after normalization, and consequently the (almost necessarily non integer) rank value at which the Pareto 80% threshold $r_P$ is reached. Such a value is given in Table 2. It is obvious that $r_P > N/5 = 4.4$.

**PLEASE PLACE FIGURE 2 ABOUT HERE**

**PLEASE PLACE TABLE 2 ABOUT HERE**

The statistical characteristics of the financial gains, as obtained, e.g., from Wessa (2018), are given in Table 2.

**PLEASE PLACE TABLE 3 ABOUT HERE**

**PLEASE PLACE FIGURE 3 ABOUT HERE**

The various concentration indices are given in Table 3.

For visualization purpose, Fig. 3 displays the yearly evolution of concentration and financial indices for the financial gains of bicycle teams in Tour de France in recent given years. Such a graphic representation of the indices trends demonstrates that they are mutually consistent.

# 4. DISCUSSION AND INTERPRETATION

A few comments are in order. First, it is easily observed that the 2015 CCDC differs from the others; it has a smaller curvature; this seems to be due to the variation in the total money prize every year. Yet, notice the the prize money went much up in 2016, but is now going mildly down. The effect on γ and $r_P$ is different. The former has a large dip in 2016, the latter has its weakest value in 2015. These differences will be further emphasized when discussing the financial indices.

Second, from a strictly numerical point of view, the γ exponent (≈1) is reminiscent of Zipf's finding about the "least effort law", also understood as an equilibrium process (Zipf, 1949). However, more modernly, it can be understood as resulting from a "self-organizing process" of complex systems, in fact, as recently discussed in a set of papers about soccer team and country ranking (Ausloos, 2014; Ausloos et al., 2014a, 2014b). The UEFA and FIFA ranking rules lead to a dissipative structure process (Prigogine and Nicolis, 1967), which ends in a stable "dissipative structure" characterized by an "equilibrium exponent" ≈1.

Third, the breakdown of Pareto law is unusual. The rule has been verified to hold elsewhere in relation to sports ranking systems (Deng et al., 2012). It is found that 20% of players possess approximately 80% of the scores or prize money of the whole system in various sports (Deng et al., 2012). However, this observation seems to have been only made for individual competitions, - and in presence of an exponential decay of the gain distribution at high rank. The present results suggests to consider further investigations on the matter.

Remarkably, this is different from a Matthew like effect, in which "the winner takes all". Let it be observed that cycling is a different matter: even though it seems that a cyclist race is won by only one individual, it is well known that this results from a team strategy and activity (Albert, 1991; Hoenigman, Bradley and Lim, 2004) as usually recognized by the winner at interviews.

*Hirschman-Herfindahl Index*

The Herfindahl-Hirschman index (HHI) is usually applied to describe company sizes (which measure the concentration) with respect to the entire market. Adapted to the case of financial gains of teams, HHI is an interesting indicator of the amount of competition. From an industry competition point of view, a HHI index below 0.01 indicates a highly competitive index (from a portfolio point of view, a low HHI index implies a very diversified portfolio). The higher the value of HHI, the smaller the number of teams with a large value of gains or the weaker the competition. The Hirschman-Herfindahl Index ranges from 0.1432 (in 2015) down to 0.1128 (in 2016), - an about 25% drop, which is very large, but not easily seen on Fig. 3. Such a large HHI value suggests that no team highly dominates others, but competition becomes fiercer. The so called Normalised Herfindahl-Hirschman index, defined as

$$HHI^* = [(HHI - 1/N)]/[1-1/N], \qquad (6)$$

varies from $\approx 0.1024$ to $\approx 0.0706$; it presents the largest drop of the whole set of coefficients, between 2015 and 2016, that is $\approx 31\%$. The HHI* has also the largest growth rate, $\approx 40\%$ of all coefficients, after 2016.

*Entropy and hierarchy*

As it has already been understood, the entropy measure should be negatively correlated to the concentration level. In other words, the closer the entropy value to 0, the higher the concentration, and the closer the entropy to its maximum *ln(N)* value, here *ln(22)=3.091*, the lower the concentration. In 2015, the entropy measure is ≈ 2.43 while in 2016 its value was ≈ 2.59, which indicates a slight decrease of concentration. One easily calculates (*Max. Entropy - Entropy*) and observes that this value presents a dip ≈ - 0.276, in 2016. Again, it can be concluded that the financial gains are "weakly concentrated".

The same conclusion can be drawn based on the Theil entropy measure: in 2015, the Theil index is ≈ 0.66 and in 2016 falls to ≈ 0.50, and regrowths to ≈ 0.60. Such data implies that there is a marked decrease in concentration at first, followed by a trend toward higher concentration, which can be understood on the basis of an increasing strength of the "best team" (SKY) in the competition, as measured at the end of the race.

The coefficient of variation goes from ≈ 1.466 down to ≈ 1.217, later on going back up to ≈1.438. The decay rate ≈ - 17 % is more recently followed by a ≈+18 % growth rate.

Based on these values, it can be said that the coefficient of variation is the more volatile measure (it can be deduced from Table 2 that CV has the largest standard deviation) of all the measures used in the research for the analysed period.

As it has been defined earlier, a high coefficient of variation implies a great dispersion around the mean. In the present case, the CV is not very large, ≈1, interestingly pointing, as also the high kurtosis value does, see Table 2, to a quite peaked distribution near the mean. For completeness, the skewness value points to a long upper tail (in a continuum approximation), again indicating the "financial gain superiority" of the "best team" and a weak concentration system.

*Pietra-Hoover, Gini wealth inequality distribution*

The Pietra-Hoover inequality index (PHI) is a peculiar measure (Delalić, et al., 2018) indicating the proportion of the total variable value which should be transferred from the value area above the arithmetic mean to the value area below the arithmetic mean such that a uniform distribution can be achieved. The value of this index can be found from the Lorenz curve: the value is obtained from the greatest "vertical distance" between the Lorenz curve and the uniform distribution line. High values of the index obviously represent a high inequality level since a greater redistribution of values is required in order to achieve equality; vice-versa, lower values of the index represent a lower inequality level. In the present case, PHI goes from $\approx 0.455$ down to $\approx 0.400$ and slowly back up to $\approx 0.410$, thus with respective $\approx -0.12$ and $\approx 0.023$ rates. This indicates that the "financial ranking" of teams is in a quasi steady state, - there is no money redistribution over the years, confirming the observation of quasi equilibrium, seen from the RSL.

Recall that in contrast to the PHI, Gi measures a ratio between surfaces. The values of the Gini coefficient range from 0.58 to 0.52, the dip here occurring in 2017. These rather low values indicate a rather low concentration. Only a subjective conclusion can be reached from the near 0.5 value: the gain distribution is not too uniform but not far from uniform, - in a statistical sense!

*Financial Concentration*

In contrast to Gi, which measures "inequality", the Rosenbluth Index is more related to a concentration measure. Besides, in contrast to the HHI which emphasizes the role of the best teams (at low ranks), RI emphasizes the role of the "worst teams". The low RI values (near ≈ 0.10) indicate that the "worst teams" are almost all equal. This is indeed in agreement with the long « flat tail », that is, the exponent of the RSL.

Moreover, based on the Rosenbluth Index values, found in Table 3, it can be said that the RI is the less volatile, when the presently less sensitive, measure: it has the lowest standard deviation ≈ 0.0062, lower than the HHI (≈ 0.013) or HHI* (≈ 0.014), of all the measures used in this research for the investigated races. Notice that like Gi, the largest RI dip occurs in 2017.

Finally, based on the data presented in Table 3 and Fig. 3, it can be concluded that the "gain concentration " had been slightly decreasing with a (≈ 0.109) rate until 2017, but has grown again in 2018, with a ≈ 0.065 rate.

Notice that the Squared Coefficient of Variation and the Normalised Herfindahl-Hirschman index, the *Max. Entropy – Entropy* and the Theil Index, as well as the Gini and concentration index, are pairs of indices which these have necessarily the same relative decay or growth rate values.

### 5. CONCLUSIONS

Several questions have been considered in this research. First, the "behavior" of the financial gains of teams in recent (male) TdFs is investigated. It is searched whether there is a scaling law as often found in other socio-economic cases. It is found that such a RSL exists and is characterized by a simple exponent reminding of the occurrence of a steady state in dissipative structure, - whatever the year of the race. However, the Pareto sparsity law is not obeyed: 80 % of the financial prizes are distributed in many more than the 4 best teams as should be theoretically expected if the Pareto's law holds true. In so doing, it is understood that the financial gain distribution at the end of the race is very particular. The official distribution rules are such that the financial gain by a team is a complex sum of inputs by the various riders. The Pareto's law breakdown might be due to this musketeer-like aspect of such long races: "one for all, all for one".

In the subsequent analysis, we have searched for various financial indices in order to answer a question about the evolution of the concentration of gains. Indeed, it is commonly accepted that several teams are better than others and would gain more money than others at the end of the race. Contrary to expectations, it is found that the financial concentration, as discussed in financial studies, is not strong at all. An explanatory conjecture stems from what is also "common expectation by observers": several teams might join together at some time and for short period of time, a couple of stages, in order to impede a too easy win by the "best rider" in a commonly admitted favourite team (Albert, 1991; Hoenigman et al., 2016). By extension, this brings some impetus to research on whether people perform better in groups only "when members of the group are individually identified and responsible" and conversely, whether people perform worse in groups when they "are not publicly identified or rewarded" (Baumeister et al. 2016; Zhang et al., 2018). In other words, how individual responsibility, obligingness, and skill contribute to group success.

Within this framework, 8 financial indices connected to the question of the concentration have been tested and compared with each other in order to put into evidence their respective meaning concerning the various parts of the gain distribution, in relation to team ranking.

In conclusion, let it be emphasized that the report presents a rare type of data, with some "universality feature potential". Several features are rather unexpected, - even confronting the sporting common fan expectations. *A posteriori*, those might be understood if one reconsiders the gain distribution rules and the riders (or teams) strategies. At this time, there is no theoretical explanation of the value of the RSL exponent. Let it be observed, through the 2019 results, see Appendix, that a « king effect » (Laherrere and Sornette, 1998) markedly exists, without a « vice-roy effect » (Cerqueti and Ausloos, 2015 ; Ausloos and Cerqueti, 2016a). Notice also that the decay parameter for (and from) the exponential fit can be hardly guessed in advance.

Alas, a complex model is out of the aim of this report and findings, yet these are suggesting to look at such similar data in other equivalent types of races, including related female team competitions, and maybe in other sports in which the team ranking derives from different inputs by the teammates. An opened question, but data has to be made available by organizers, is whether prize money distribution rules in different sportively similar races lead to observe universal features, and if not, why they vary. We like to think that financial gains measure performance, but do they ?

# APPENDIX

PLEASE PLACE FIGURE 4 ABOUT HERE

**The 2019 Tour de France case**

Due to the delay in the peer review process, the 2019 Tour de France gains are now available. Thus the RSL may be examined results can be examined. The pertinent display is found in Fig. 4. A power law fir leads to the value of the exponent : -1.2, similar to that found for the previous years.

In so doing, it can be pointed out here that the exponent seems to be mainly originating from the so called « king effect » (Laherrere and Sornette, 1998); the tail of the RSL rather looks like an exponential. Inded, if one arbitrarily removes the 1st team gains, a better fit than the power law is the exponential. The decay parameter is about equal to 7.58. Notice that the vice-roy effect (Cerqueti and Ausloos, 2015; Ausloos and Cerqueti, 2016a) seems very weak.

**Endnotes**

[1] The 80–20 "Pareto rule" states (Pareto and Page, 1971) that, in many cases, roughly 80% of the effects come from 20% of the cause.

[2] There are team competitions, like in "against the clock" or "team time trial" races.

[3] Professional cycling is one of the oldest professional sports (Desbordes, 2006, 2008).

[4] In classical finance studies, the upper limit of the sum in Eq. (1) is conventionally limited to the supposedly most important (up to 30) firms on the market, but there is no criticism to provide if Eq. (1) used if *N<30*.

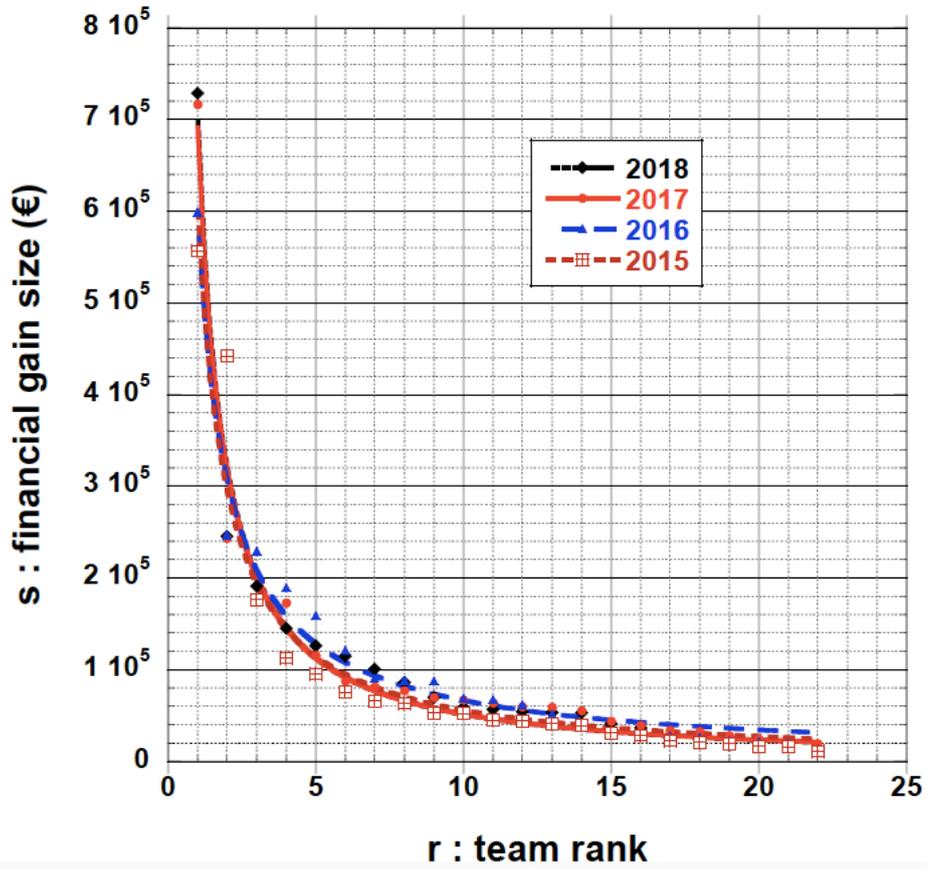

Fig. 1 Rank ($r$)-size ($s$) law of financial gains in EUR for bicycle teams in Tour de France in given years.

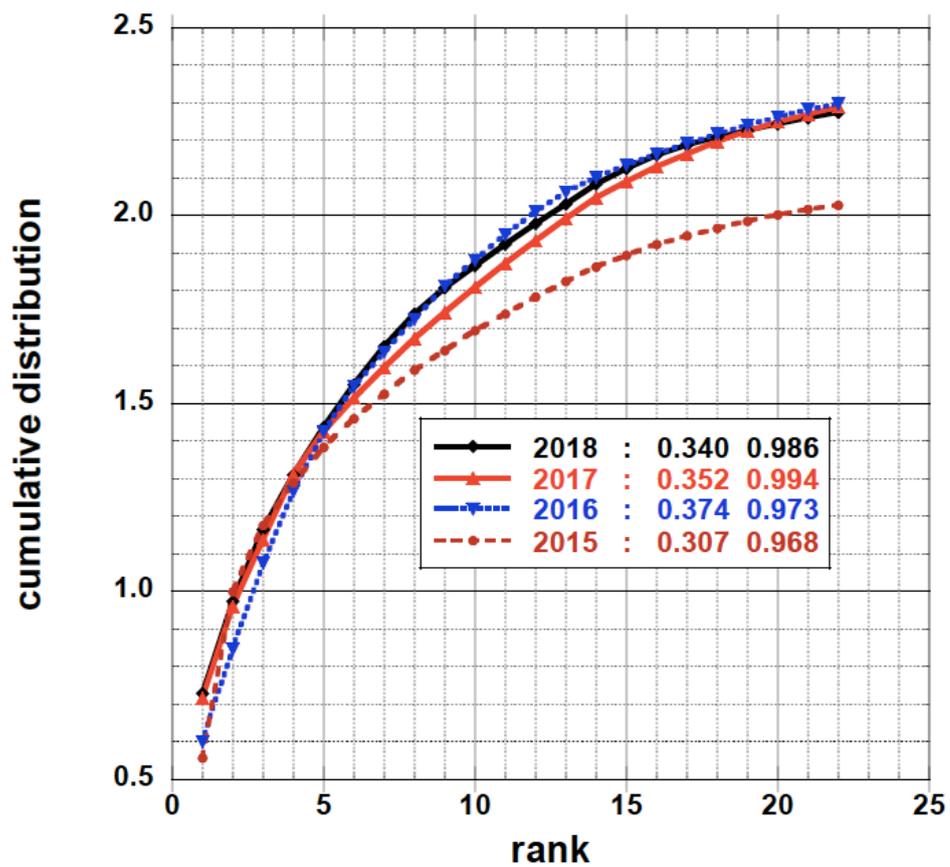

Fig. 2 Cumulative concentration distribution curve of financial gains in million euros for bicycle teams in Tour de France in given years. The inset gives the power law exponent (obtained through a Levenberg-Marquardt algorithm) and the $R^2$ value of the fit.

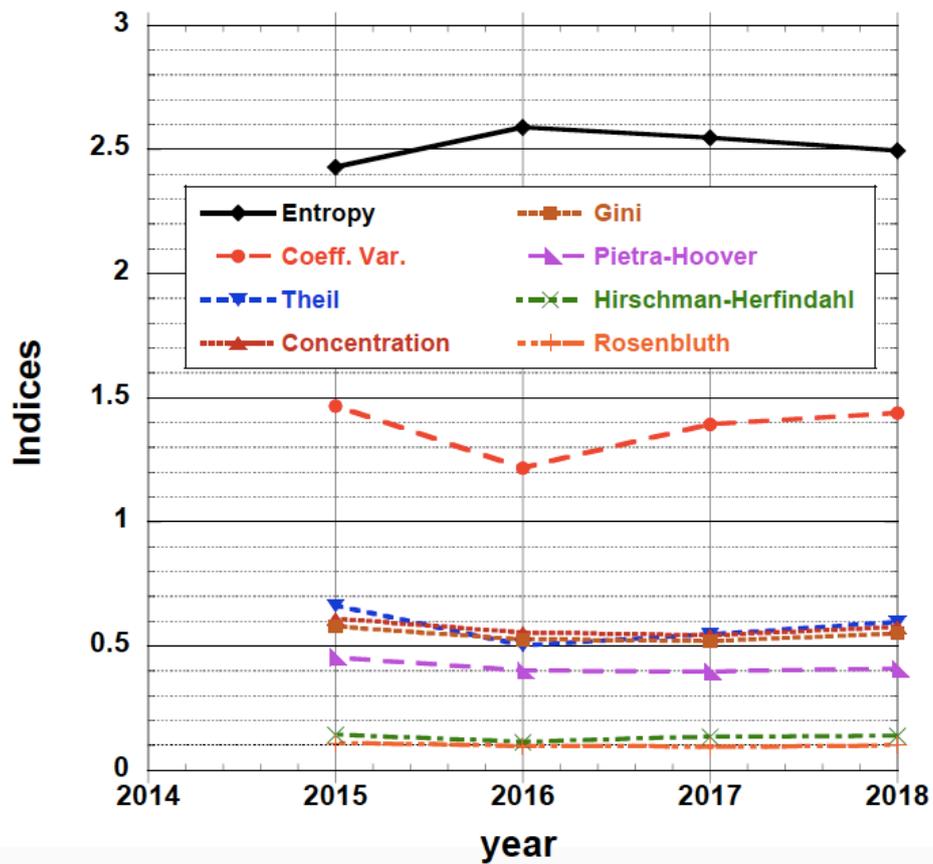

Fig. 3 Yearly evolution of concentration and financial indices for the financial gains of bicycle teams in Tour de France in recent given years.

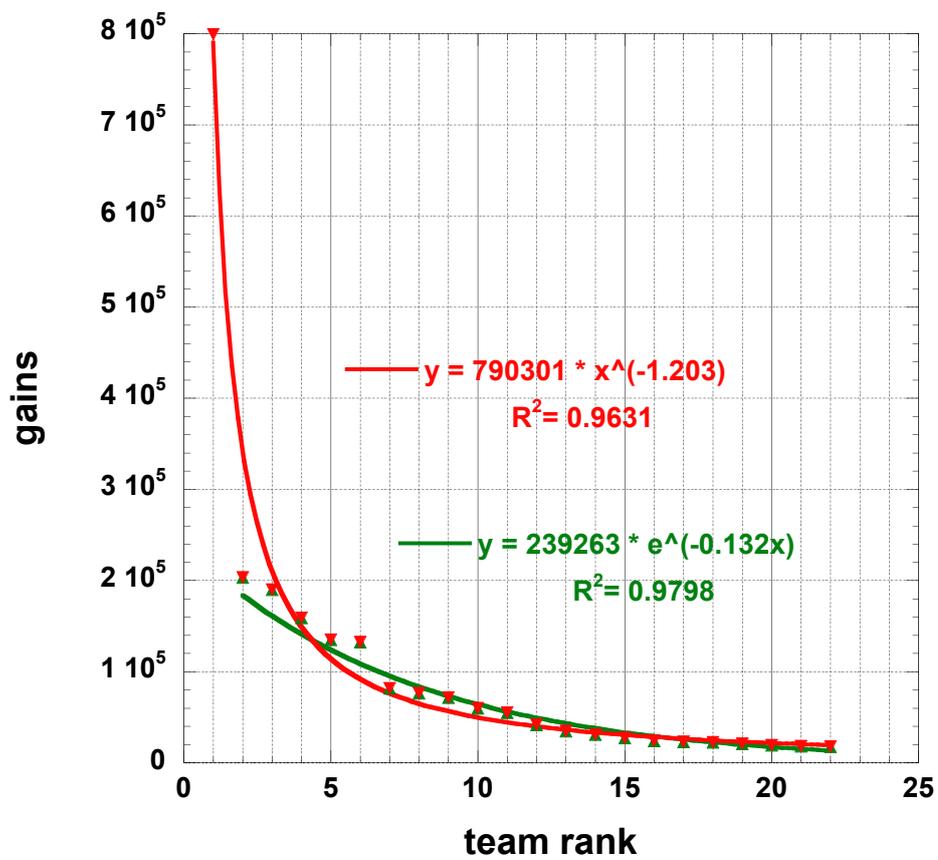

Fig. 4 Rank (*r*)-size (*s*) law of financial gains in EUR for bicycle teams having competed in 2019 Tour de France. One may distinguish a « king effect ».

| year | α x 10$^{(-5)}$ | γ | R$^2$ | r$_P$ |
|------|-----------------|---|-------|-------|
|      |                 |   |       |       |
| 2015 | 6.00+/-0.35     | 1.03+/-0.07 | 0.933 | 8.65 |
| 2016 | 5.90+/-0.19     | 0.95+/-0.03 | 0.978 | 9.35 |
| 2017 | 6.92 +/-0.21    | 1.13+/-0.04 | 0.980 | 10.34 |
| 2018 | 7.05 +/-0.21    | 1.14+/-0.04 | 0.981 | 9.2 |

Table 1. Best-fit parameters of the Zipf's law, Eq(5). The standard error at 95% for the two parameters are reported with the subsequent R$^2$; the "Pareto rank" r$_P$ is given (see text for definition).

| year | Min. | Max. | Total | Mean | Std. Dev. | Skewness | Kurtosis |
|------|------|------|-------|------|-----------|----------|----------|
|      |      |      |       |      |           |          |          |
| 2018 | 14 420 | 728 630 | 2 272 250 | 1.033 | 1.5208 | 3.3720 | 11.339 |
| 2017 | 19 230 | 716 590 | 2 287 650 | 1.040 | 1.4820 | 3.4367 | 11.638 |
| 2016 | 14 100 | 599 240 | 2 295 850 | 1.044 | 1.3002 | 2.7072 | 7.6602 |
| 2015 | 10 940 | 556 630 | 2 027 650 | 0.922 | 1.3833 | 2.5755 | 5.4239 |

Table 2. Statistical characteristics of the financial gain distributions of the 22 teams at the end of the Tour de France for given race years; Min. , Max. , and Total are in EUR; Mean and Std. Dev. are in 10$^5$ EUR.

| year | 2015 | 2016 | 2017 | 2018 |
|---|---|---|---|---|
| Hirschman-Herfindahl Index | 0.14319 | 0.11281 | 0.13359 | 0.13948 |
| Entropy | 2.42815 | 2.58900 | 2.54552 | 2.49462 |
| Theil Index | 0.66290 | 0.50204 | 0.54553 | 0.59642 |
| Coefficient of Variation | 1.46638 | 1.21728 | 1.39244 | 1.43826 |
| Pietra-Hoover Index | 0.45496 | 0.40030 | 0.39612 | 0.40948 |
| Gini Index | 0.58079 | 0.52707 | 0.51757 | 0.55123 |
| Rosenbluth Index | 0.10843 | 0.09611 | 0.09422 | 0.10129 |
| Concentration Index | 0.60845 | 0.55217 | 0.54222 | 0.57748 |

Table 3. Financial inequality and concentration measure indices for the 22 cycling team financial gain distributions at the end of the Tour de France in given years.